# Semiconductor Nanowire Light Emitting Diodes Grown on Metal: A Direction towards Large Scale Fabrication of Nanowire Devices

ATM Golam Sarwar[1], Santino D Carnevale[1], Fan Yang[2], Thomas F Kent[2], John J Jamison[2], David W McComb[2], and Roberto C Myers[1,2*]

[1]Department of Electrical and Computer Engineering, The Ohio State University, Columbus, Ohio 43210, USA.

[2]Department of Materials Science and Engineering, The Ohio State University, Columbus, Ohio 43210, USA

*E-mail: myers.1079@osu.edu.

**Bottom up nanowires are attractive for realizing semiconductor devices with extreme heterostructures because strain relaxation through the nanowire sidewalls[1][2] allows the combination of highly lattice mismatched materials without creating dislocations. The resulting nanowires are used to fabricate light emitting diodes (LEDs), lasers, solar cells and sensors[3][4][5][6]. However, expensive single crystalline substrates are commonly used as substrates for nanowire heterostructures as well as for epitaxial devices, which limits the manufacturability of nanowire devices. Here, we demonstrate nanowire LEDs directly grown and electrically integrated on metal. Optical and structural measurements reveal high-quality, vertically-aligned GaN nanowires on molybdenum and titanium films. Transmission electron microscopy confirms the composition variation in the polarization-graded AlGaN nanowire LEDs. Blue to green electroluminescence is observed from InGaN quantum well active regions, while GaN active regions exhibit ultraviolet emission. These results demonstrate a pathway for large-scale fabrication of solid state lighting and optoelectronics on metal foils or sheets.**

The III-Nitride semiconductor family has a wide range of technologically important applications, especially in visible and ultraviolet light emitting diodes (LEDs) [7][8] and lasers. High quality GaN thin films are commonly grown on single crystalline sapphire substrates due to the good epitaxial relationship [9]. Since the first synthesis of GaN on sapphire major technological hurdles have been overcome to realize the commercialization of III-N based LEDs on sapphire such as realizing p-type doping [10], mitigating electron overflow using wide bandgap AlGaN electron blocking layer [7], and growth of high quality InGaN and AlGaN quantum wells [11][12].

Recently a number of research groups have been actively working on III-N nanowires, and significant progress has been reported on the growth method/mechanisms [13][14][15][16][17][18][19][20][21][22][23][24][25][26][27][28][29]. Some commonly used synthesis methods include catalyst based nanowires, self-assembled catalyst free nanowires, and nanowires on pre-patterned substrates or selective area growth. Materials synthesized using catalysts suffer from high concentrations of unwanted impurities and selective area growth requires costly and complicated processing steps. On the other hand catalyst free GaN nanowires are free of these disadvantages. Single crystalline Si substrates are commonly used to produce self-assembled catalyst free GaN nanowires; Si has the advantage of also serving as a contact for electronic and optoelectronic devices. However, one of the major limiting factors in the realization of large scale fabrication is the single crystal Si substrate. It also limits performance as it absorbs light in optoelectronic devices. Thankfully, GaN nanowires have been shown to grow on optically transparent amorphous $SiO_x$ [30][31] and glass substrates [32]. These results are proof that GaN nanowires do not need a global epitaxial relationship with the substrate and can be grown on largely scalable substrates.

Unfortunately, the aforementioned substrates are electrically insulating and cannot be used for devices requiring electrical conduction. Therefore, realization of high quality nanowire growth on conductive layers is necessary. One option is to grow nanowires directly on smooth metal films deposited on amorphous substrates (glass, $SiO_x$ etc.) or on metal foils. The latter approach opens the possibility of large scale fabrication, e.g. roll to roll fabrication process, using metalorganic chemical vapor deposition (MOCVD) of flexible electronic and optoelectronic devices. Furthermore, nanowires directly grown on very rough metal sheets might be useful in applications where direct electrical contact formation is not necessary. One example of such application is photo-chemical or photo-electrochemical $H_2$ generation using Ga(In)N nanowires [33].

Though direct growth of semiconductor nanowires on metals was demonstrated in several material systems, such as Si, Ge, CdS, and ZnO [34][35][36][37], it was only very recently reported by Wölz et al. for GaN nanowires grown on sputtered Titanium (Ti) films [38]. However, operating III-Nitride nanowire devices electrically integrated and directly grown on metal were not yet demonstrated to our knowledge.

Here we report the direct growth and electrical integration of III-Nitride nanowire LEDs on metal. First, the growth of GaN nanowires on thin Mo and Ti films on Si wafers is demonstrated. Photoluminescence (PL) measurements reveal similar optical quality for nanowires grown on metal compared to those grown on Si wafers. Room temperature band edge PL emission at 363 nm with a 10 nm full width at half maximum (FWHM) is observed while below band gap defect related PL is absent. Next, nanowire LEDs composed of polarization-graded AlGaN nanowires on thin Mo films are fabricated with a variety of InGaN and GaN active regions. The electroluminescence (EL) from InGaN nanowire LEDs is varied from blue to green (450 to 565 nm) with a peak efficiency at 230 A/cm$^2$ and 3% efficiency droop at 500 A/cm$^2$. Modeling of the external quantum efficiency (EQE) using the ABC model reveals a maximum internal quantum efficiency (IQE) of ~47% which is comparable to the reported typical IQE values for visible GaN based nanowire LEDs on Si. Finally, nanowire LEDs grown on Mo films containing GaN active regions emit at ultraviolet (385nm) wavelengths. The choice of Mo metal film is governed by the fact that our nanowire LEDs pose a p-down orientation due to the use of polarization induced doping and Mo is expected to make good electrical contact with p-type GaN due to its relatively large work function. Additionally it is a safe material for molecular beam epitaxy (MBE) environment due to its extreme low vapor pressure at growth temperatures.

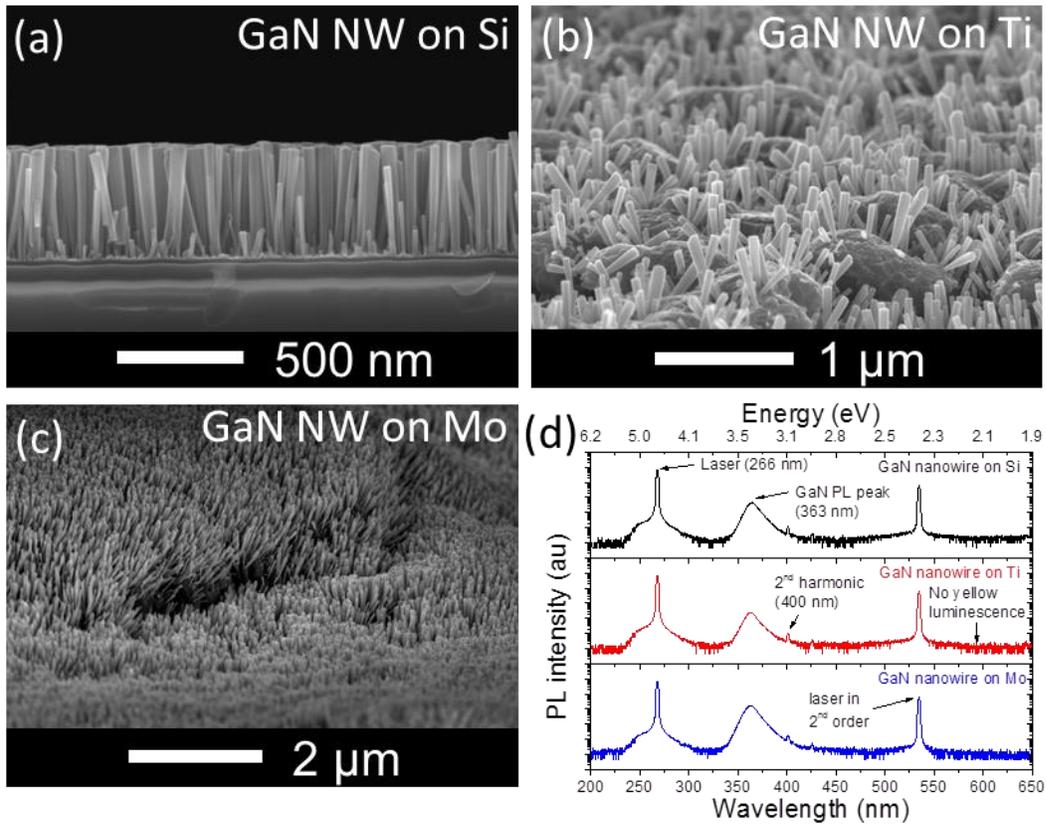

**Figure 1: Growing semiconductor nanowires on metal films. Scanning electron microscopy (SEM) images of GaN nanowires grown on (a) Si, (b) Ti, and (c) Mo. (d) Room temperature photoluminescence (PL) measurements of GaN nanowires grown on Si (black line), Ti (red line), and Mo (blue line).**

Self-assembled catalyst free GaN nanowires are grown directly on Ti and Mo films using plasma assisted molecular beam epitaxy (PAMBE). To ensure identical growth conditions, one quarter of the wafer is coated with Ti, another quarter is coated with Mo and the rest of the wafer is left as bare Si. The metal coated Si wafer is transferred to the MBE growth chamber and native oxide desorption is performed *in-situ* by heating the substrate to $1000^{\circ}$ C for one minute with a ramp rate of $50^{\circ}$ C/min under a vacuum of $\sim 7 \times 10^{-11}$ torr. Figure 1(a), (b) and (c) show the scanning electron microscopy (SEM) images of the nanowires grown on Si, Ti and Mo, respectively. As expected, the nanowires on Si exhibit vertical orientation. Statistical analysis of the SEM images reveals that density and diameter of the nanowires on Si are $209 \pm 12$ $\mu m^{-2}$ and $34.75 \pm 8.56$ nm, respectively. It is found that nanowires grown on Ti film exhibit relatively lower density and larger diameter ($54.77 \pm 7.76$ nm) while nanowires grown on Mo film exhibit similar density and diameter ($35.75 \pm 5.81$ nm) compared to the nanowires grown on Si. In both metals, delamination of the film from the Si wafer is observed, as shown in Fig. 1 (b) and (c). We speculate that this occurs due to the large thermal expansion mismatch between Si ($2.6\text{-}3.3 \times 10^{-6}$ $K^{-1}$) and metal films (Mo: $4.8\text{-}5.1 \times 10^{-6}$ $K^{-1}$ and Ti: $8.4\text{-}8.6 \times 10^{-6}$ $K^{-1}$). Moreover, the high temperature ($1000^{\circ}$ C) *in-situ* oxide desorption step might also contribute to the delamination of metal films. Usage of small ramp rate and skipping the *in-situ* oxide desorption step (which is not necessary if solely grown on metal) is found to help reduce the delamination problem and used for later samples.

Figure 1 (d) shows the room temperature photoluminescence (PL) emission spectra from nanowires grown on Si (black), Ti (red), and Mo (blue). Nanowires grown on all three substrates show only a single PL peak emission at 363 nm (~3.423 eV) with FWHM of 10 nm, which corresponds to band to band carrier recombination in GaN. Additional, sub-band gap features are not observed over the measured spectral range (300 to 650 nm). PAMBE grown GaN nanowires on Si are reported to be nearly dislocation free and do not exhibit defect-mediated yellow luminescence which is very common in GaN thin films. The absence of yellow luminesce from nanowires grown on Ti or Mo films indicates that nanowires grown on metal films possess similar quality as nanowires grown on single crystalline Si wafers.

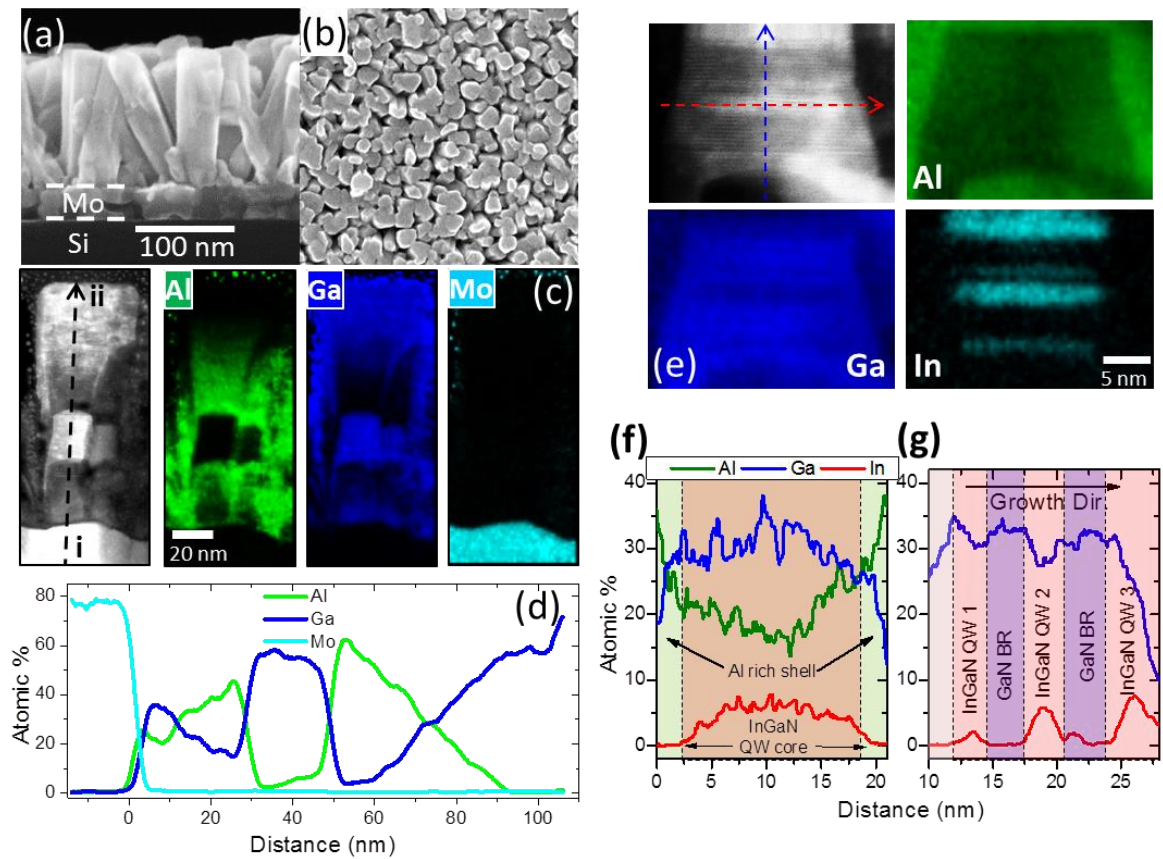

**Figure 2: Chemical mapping of the compositional variation in nanowire heterostructures grown on metal. (a) Cross-section and (b) plan view SEM images of graded AlGaN nanowire LEDs grown on**

**Mo films deposited on Si substrate. (c) High angle annular dark field (HAADF) scanning transmission electron microscopy (STEM) image along with energy dispersive x-ray spectroscopy (EDXS) chemical composition maps showing back and forth composition grading. (d) EDXS line scan along the axis (i-ii in Fig. 2(c)) of the nanowire heterostructure showing approximately linear variation of Al and Ga composition. Mo signal can be observed from beneath the nanowire. (e) HAADF STEM image of the active region of a single nanowire. Ga, Al and In composition maps are also shown. (f) EDXS line scan along the radial direction through the 2$^{nd}$ quantum well (QW) (red dashed line in Fig. 2(e)) showing InGaN QW wrapped in an Al rich AlGaN shell. (g) EDXS line scan along the axis of the active region (blue dashed line in Fig. 2(e)) showing three InGaN QWs with increasing In composition in the growth direction.**

After successfully growing high quality GaN nanowires on metal films, we grow and characterize nanowire LEDs on metal with both visible and near ultraviolet emitting active regions. The design of the nanowire LED heterostructure takes advantage of the intrinsic polarization of III-N materials. When compositionally graded, the polarization creates 3D distributed bound charge. Here, we nucleate Mg-doped GaN nanowires on Mo films using PAMBE at 720$^o$ C. A previous study showed that catalyst free PAMBE grown GaN nanowires grow preferably in the $(000\bar{1})$ crystallographic direction (N-face) [39]. Following nucleation, the composition of the nanowire is linearly graded using a shutter pulsing method [40] from GaN to AlN over 50 nm. This creates $6.5\times10^{18}$ cm$^{-3}$ negative bound polarization charge, resulting in p-type conductivity. An active region is comprised of 3× InGaN/GaN quantum wells deposited at 625$^o$ C. Next, a 100 nm AlN to GaN linearly graded layer is deposited to induce polarization doped n-type layer. The bottom (p-type) and top (n-type) graded regions are doped with Mg and Si, respectively. Interested readers are directed to ref. [41] for details on polarization induced doping in graded AlGaN nanowires and its limitations.

Fig. 2 (a) and (b) show the cross-section and plan-view SEM images of the nanowire LED heterostructure, respectively. Nearly vertically aligned high density nanowires are observed on the Mo film. Fig. 2(c) shows high angle annular dark field (HAADF) scanning transmission electron microscopy STEM image and energy dispersive x-ray spectroscopy (EDXS) chemical composition (Al, Ga and Mo) maps in the nanowire heterostructure. Chemical composition maps show a maximum Ga concentration at the base and top of the nanowire and maximum Al concentration at the middle of the nanowire confirming the back and forth compositional grading in the vertical direction. Shown in Fig. 2(d), the EDXS line scan along the growth direction (i-ii in Fig. 2(c)) demonstrates the desired linear change in Ga and Al composition in the n- and p-side as designed. The Ga rich section at the center of the nanowire (Fig. 2(c)) corresponds to the active region. We do not see detectable In signal from the active region, which is likely due to the overlapping of multiple nanowires in this sample prepared using FIB. However, In signal is observed from a single nanowire that is dispersed on TEM grid. Fig. 2(e) shows a HAADF STEM image of the active region of a nanowire LED and corresponding chemical composition maps (Ga, Al and In). An EDXS line scan along the radial direction (red line in Fig. 2(e)) through an InGaN quantum well (Fig. 2(f)) shows a defined InGaN core region wrapped with an Al-rich AlGaN shell. The spontaneous formation of an Al-rich AlGaN shell region is the result of low adatom mobility of Al compared to Ga and In at growth temperatures. The Al adatoms impinging on the side walls incorporate where they land before they can diffuse to the top of the nanowire. On the other hand, In and Ga adatoms have a higher mobility (compared to Al) and can diffuse towards the top of the nanowire and incorporate there. A line scan along the axis of the active region (blue line in Fig. 2(e)) shows the formation of three InGaN quantum wells separated by GaN barriers, shown in Fig. 2(g). The In composition, and thickness of the quantum wells is found to increase in subsequent quantum wells. This type of variation in composition and thickness was observed previously in InGaN/GaN nanowire heterostructure by G. Tourbot et al. [42]. The authors attributed this variation to composition pulling effect [43]. It is worth noting that the EDXS analysis was performed without a standard and due to the complex core-shell geometry of the nanowires the composition obtained in this analysis is not absolute.

Nanowire LED devices are fabricated by depositing a 5 nm Ti / 10 nm Au semitransparent top contact directly onto the tops of the nanowires. A 300 nm Au grid is deposited on top of the semitransparent contact as a current spreading layer. An In-diffused bottom contact is formed on the Si substrate after mechanically removing the nanowires. The schematic of the contact

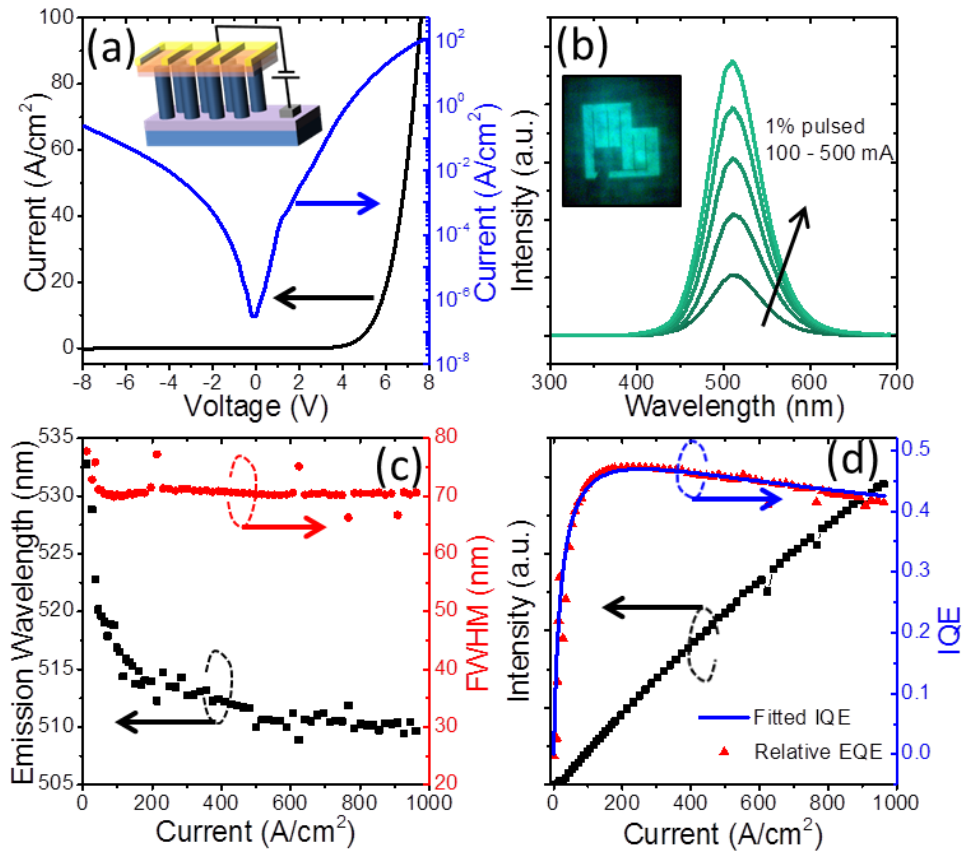

**Figure 3: Green nanowire LEDs grown on metal with InGaN quantum well active regions. (a) Current-voltage characteristic of the nanowire LED on Mo film. Inset shows the schematic of contact formation scheme. (b) Electroluminescence (EL) spectra under pulsed current excitation of an InGaN QW nanowire LED grown on Mo film emitting near green band. Inset shows an optical image of an LED device in operation. (c) Peak emission wavelength (black squares) full width at half maximum (FWHM) as a function of current density. (d) EL peak intensity (black squares) and relative EQE (red triangles) as a function of injection current density. IQE (blue line) is modeled by fitting relative EQE using ABC model.**

formation scheme is shown in the inset of Fig. 3(a). A DC current-voltage (I-V) characteristic of a nanowire LED device is shown in the Fig. 3(a). Diode behavior with a threshold voltage of ~6V is observed. EL measurements are performed on 300×300 μm$^2$ LED devices under pulsed (~1% duty cycle) current excitation to minimize any junction heating effect. Fig. 3(b) shows EL spectra from a device under excitation over the range of 100 – 500 mA in 100 mA steps. The inset of Fig. 3(b) shows an optical image of the nanowire LED device operating under DC excitation. A video of an operating nanowire LED on Mo film can be found in the supplementary information. Peak emission at ~510 nm is observed which corresponds to emission from the multiple InGaN quantum wells. Fig. 3(c) shows peak emission wavelength (black squares) and full width at half maximum (FWHM) (red circles) as a function of injection current density. Peak emission shifts from ~532 nm (at low injection) to ~510 nm (at high injection). We attribute this blue shift to a combination of two factors. First, this can be due to quantum confined Stark effect in individual QWs commonly observed in nitride based polar quantum wells. Second, this could be the result of a slight variation in In composition in QWs. Due to better electron injection compared to hole injection, emission from the third QW (close to n-side graded region) dominates at low injection. With the increase in carrier injection, emission from second and first

QW causes the blue shift due to decreased In composition (i.e. high bandgap) in these QWs. The FWHM decreases from 77.7 nm at low injection to 70.6 nm at high injection. Fig. 3(d) shows peak emission intensity (black squares) and relative external quantum efficiency (EQE) (red triangles) as a function of current density. Peak emission intensity increases linearly from low to moderate injection and shows deviation from linearity at high current density. EQE shows an increase with current density and peaks at ~230 A/cm$^2$. EQE value drops by 3% at 500 A/cm$^2$ and 12% at 1000 A/cm$^2$, compared to the peak EQE value.

The IQE of an LED can be modeled using the ABC model.

$$IQE = \frac{Bn^2}{An + Bn^2 + Cn^3}$$

Here, A is the Shockley-Read-Hall non-radiative recombination coefficient, B is the radiative recombination coefficient, C represents higher order carrier loss due Auger recombination and/or carrier overflow, and n is the carrier concentration in the active region. EQE is the product of IQE and Extraction Efficiency (EE). Assuming EE to be independent of injection current density, IQE is modeled by fitting the shape of EQE using the ABC model. Using this method we found A, B and C are $2 \times 10^8$ s$^{-1}$, $2.5 \times 10^{-10}$ cm$^3$s$^{-1}$, and $10^{-28}$ cm$^6$s$^{-1}$, respectively, for our nanowire devices, very close to the reported values for InGaN nanowire LEDs [44]. The maximum IQE is found to be ~ 47% which lies in the reported value range (40-50%) for visible nanowire LEDs[33][45].

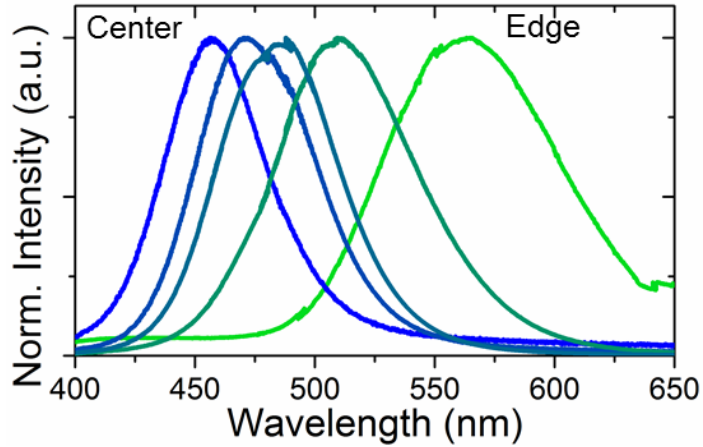

**Figure 4: Demonstration of blue to green nanowire LEDs grown on metal. A wide range of emission wavelengths is observed from different regions of a three inch wafer coated with Mo due to the temperature sensitivity of In incorporation in the InGaN active regions.**

The nanowires in this study are grown on a three inch wafer. Fig. 4 shows emission spectra of the LED devices measured from different locations of the wafer. Right-most (left-most) represents an LED device from edge (center) of the wafer. We observed a variation in the emission wavelength from LED devices across the wafer. We attribute this variation to the temperature gradient across the three-inch substrate during nanowire deposition. It is well known that InGaN composition is a strong function of temperature which results in higher In incorporation in the QWs at the edge (low temperature) than in the center (high temperature) region. Furthermore, SEM image analysis on the nanowire samples revealed smaller average diameters (31.6 nm) at the center (high temperature) of the wafer compared to the average diameter (39.9 nm) at the edge (low temperature) of the wafer. R. Armitage et al. have shown that In incorporation in nanowires increases with increasing diameter [46]. These two facts explain the blue shift observed in the EL spectra in Fig. 4 from the edge to center region.

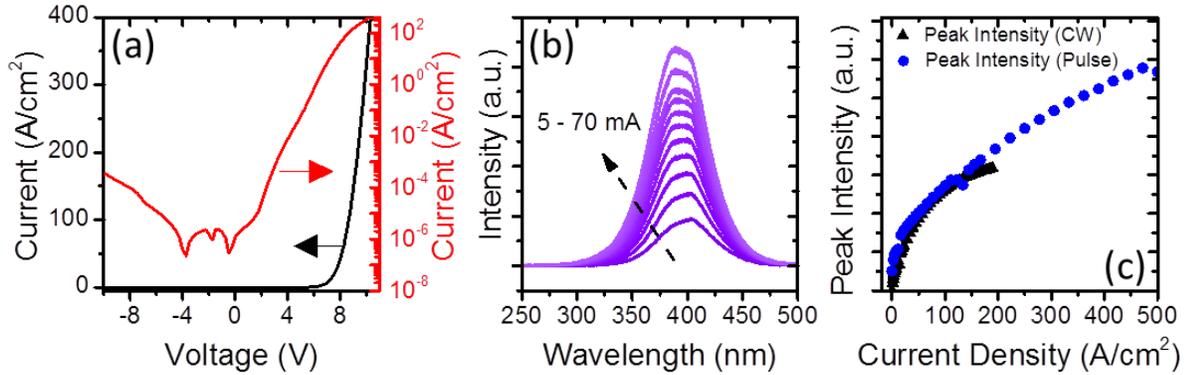

**Figure 5: Ultraviolet nanowire LEDs grown on metal. (a) Current-voltage characteristic of a GaN QW nanowire LED grown on Mo film. (b) Electroluminescence (EL) spectra under pulsed current excitation of a GaN QW nanowire LED grown on Mo film emitting near ultraviolet wavelength. (c) EL peak intensity under continuous wave, CW (black triangles) and pulsed (blue circles) current as a function of injection current density.**

We also prepared a sample replacing the InGaN active region with a 10 nm GaN quantum well. I-V characteristic of the nanowire LED is shown in Fig. 5(a). This device exhibits larger threshold voltage (~7.2 V) compared to the InGaN active region device (~6 V). This is expected due to the increased band gap of the active region. Fig. 5(b) shows the emission spectra of the LED device under DC excitation from 5 – 70 mA. The emission peaks in the near UV wavelength (~385 nm). Fig. 5(c) shows peak intensity vs current density under DC (black triangles) and pulsed (blue circles) excitation. Peak intensity increases very rapidly at very low current densities and shows immediate downwards bending.

In summary, III-Nitride based nanowires LEDs can be directly grown and electrically integrated on metals with comparable performance to those grown on Si wafers. In this study, we deposited metal on a Si wafer as a proof of concept. However, polycrystalline metal films can be deposited on any large smooth (amorphous or crystalline) substrate without degrading the film quality. Nanowires grown directly on such metal films can be used for the very large scale fabrication of electronic and optoelectronic devices. For non-contact applications, for example, photo-chemical or photo-electrochemical solar fuel generation via splitting water into hydrogen and oxygen, nanowires can be grown on rough metal plates or foils. Thus, a new pathway for large scale fabrication of nanowire-based photonics on metal is opened with the possibility to greatly reduce the cost of current solid state lighting, which requires expensive single crystalline substrates. This development may lead to widespread use of nanowire based optoelectronics.

**METHODS**

**PAMBE growth of GaN Nanowires:**

Self-assembled catalyst free GaN nanowires are grown using Veeco Gen 930 plasma assisted molecular beam epitaxy system. Mo and Ti thin films (~50 nm) are deposited using electron beam evaporation on a three inch Si (111) wafer. Nanowires are grown using the two-step dynamic growth mechanism described in [24]. In this method nanowires are nucleated at low temperature until required nanowire density is obtained followed by a high temperature growth in which already nucleated nanowires continue to grow but new nanowire nucleation does not occur. GaN nanowires are nucleated at $720^{o}$ C for 5 minutes and grown at $790^{o}$ C for 2 hours. A nitrogen partial pressure of $2\times10^{-5}$ torr is used with a plasma power of 350W.

**Structural characterization:**

SEM images were collected using an FEI/Philips Sirion SEM with a field emission (FE) source and an in-lens secondary electron detector. STEM samples were prepared with a Helios NanoLab 600 dual-beam focused ion beam (FIB). High resolution STEM imaging was carried out on an FEI image corrected Titan[3TM] G2 60-300 S/TEM equipped with a quad silicon drift detector at 300 kV.

**Photoluminescence**:

The nanowires are optically excited using the third harmonic (266 nm) of a mode-locked Ti:sapphire oscillator (Coherent Chameleon Ultra II) operating at 800 nm and 40 MHz. The samples were illuminated with an average power of 0.5 mW and focused on the sample surface through a 0.5 NA 36× reflective objective which results in a beam diameter of ~10 μm. The emission from the samples were collected through a 300 nm long pass filter and passed to a 0.5 m spectrometer (Princeton Instruments SP2500i) equipped with a UV–vis CCD (Princeton Instruments PIXIS100) and a 1200 g/mm diffraction grating.

**AKNOWLEDGEMENTS:**

This work was supported by the Army Research Office (W911NF-13-1-0329) and by the National Science Foundation CAREER award (DMR-1055164).